# Exact Solution of Dirac Equation with Charged Harmonic Oscillator in Electric Field: Bound States

**Sameer M. Ikhdair**

Physics Department, Near East University, Nicosia, North Cyprus, Turkey

sikhdair@neu.edu.tr

Tel: +903922236624; Fax: +903922236622

## Abstract

In some quantum chemical applications, the potential models are linear combination of single exactly solvable potentials. This is the case equivalent of the Stark effect for a charged harmonic oscillator (HO) in a uniform electric field $\varepsilon$ of specific strength (HO in an external dipole field). We obtain the exact s-wave solutions of the Dirac equation for some potential models which are linear combination of single exactly solvable potentials (ESPs). In the framework of the spin and pseudospin symmetric concept, we calculate the analytic energy spectrum and the corresponding two-component upper- and lower-spinors of the two Dirac particles by the Nikiforov-Uvarov (NU) method, in a closed form. The nonrelativistic limit of the solution is also studied and compared with the other works.

## 1. Introduction

The Schrödinger equation provides an insight to the fundamental quantum chemical problems. There are a number of solvable nonrelativistic quantum problems in which all the energy eigenvalues and wave functions are explicitly known from different operator methods [1] and analytical procedures [2] specially developed to solve the desired wave equation. This solution can be done by using the supersymmetry (SUSY) [3,4], the Nikiforov-Uvarov (NU) method [5], the asymptotic iteration method (AIM) [6], the exact quantization rule (EQR) [7] and the tridiagonal J-matrix method (TJM) [8], etc.

The electron confinement in harmonic oscillator (HO) potential exposed to n external electric field is one of the quantum chemical applications. Indeed, this is well known as charged HO in a uniform electric field or an HO in an external dipole field. Moreover, such model could be also used in the measurement of the relative photo ionization cross section of Rb in the presence of various strengths of external electric fields [9]. The model potential also makes specific predictions about the spacing as a function of applied field and used in the calculation of the energy levels of ammonia in strong electric field [10]. The SUSY and shape invariance methods have been used to determine ESPs are extended to obtain the energy eigenvalues and their generalized partner potentials [11]. It demands the existence of the Witten superpotential $W(x)$ [12] associated with the ESPs in order to find the Witten superpotential for the combined potential. It is an ansatz used as particular solution of the involved Riccati equation [13]. The generalized eigensolutions of some important ESPs in one-dimension have been studied [11]. Further, the determination of the vibration spectra in some molecular systems is found essential in chemical study. The Morse [14], Hulthén [15,16] and Kratzer [17] potentials are models used to study diatomic molecules. The modification on the spectrum energy due to the influence of electric field on a quantum particle of mass $m$ and charge $q$ confined by HO potential was studied with a disclination [18,19].

Besides, the spherical relativistic HO with spin symmetry has been studied [20]. The Dirac Hamiltonian with scalar $S(r)$ and vector $V(r)$ potentials quadratic in space coordinates [21] has been used to find an HO like second order equation. This can also be solved analytically for Klein-Gordon (KG) equation with equally mixed scalar and vector potentials $S(r) = \pm V(r)$ (the sum potential $\Sigma(r) = 0$ and the difference potential $\Delta(r) = 0$) [22]. Recently, the triaxial, axial and spherical HO for the case $\Delta(r) = 0$ has been solved and applied to the study of anti-nucleons embedded in nuclei [23-25]. The case $\Sigma(r) = 0$ is particularly relevant in nuclear physics since it is a necessary condition for the pseudospin symmetry in nuclei [26,27]. The bound state solution of the spin-$1/2$ particles in Dirac equation with HO have been obtained by letting either $\Sigma(r)$ or $\Delta(r)$ equal to zero [28]. The perturbative breaking of pseudospin symmetry induced by a tensor potential [29] has been studied despite the condition $\Sigma(r) = 0$ or $d\Sigma(r)/dr = 0$ can not be realized in nuclei [30]. The correlation between the pseudospin splitting and the shape of the HO potential, namely the HO frequency and the distance of well-bottom deviation from the center studied in [31]. The relativistic HO in $1+1$ dimensions, i.e., including a linear potential and quadratic scalar and vector potentials with equal or opposite signs has been solved [32]. The solutions found for zero pseudoscalar potential are related to the spin and pseudospin symmetry of the Dirac equation in $3+1$ dimensions. The Dirac equation with HO scalar and vector potentials along with the tensor potential as a sum of linear and Coulomb-like potentials has been studied [33]. It was found that the tensor potential preserves the form of the HO potential and generates spin-orbit terms. The bound states of a new ring-shaped equal mixture of vector and scalar HO for spin-$1/2$ Dirac particles were studied [34]. The bound state solutions of the relativistic pseudoharmonic potential have been studied using the Nikiforov-Uvarov method [35].

Our aim is to obtain the exact s-wave Dirac bound state energies and the upper- and lower-spinor wave functions in HO potential influenced by a uniform electric field. Further, we investigate the modification on the spectrum energy of a quantum particle influenced by a uniform electric field in the radial direction in the presence of the spin symmetry $S(r) \approx V(r)$ and pseudospin symmetry $S(r) \approx -V(r)$ cases in the framework of the NU method [5,14-16,36-39]. These solutions are reduced to the spinless KG and Schrödinger limits when $S(r) = +V(r)$ and $S(r) = -V(r)$ corresponding to exact spin $\Delta(r) = 0$ and pseudospin $\Sigma(r) = 0$ symmetry limitations [23-25,40-45]. The performance of any method applied to the Dirac equation depends on the selected representation of this equation and the mathematical structure of the resulting equation, in which there are conditions under which it may be reduced to a Schrödinger equation [46-48].

In this paper, Section 2 briefly introduces Dirac formalisms. The Dirac bound state energies of a particle confined by an ESPs consisting of combined harmonic oscillator-plus-linear (HpL) potential model in the presence of spin and pseudospin symmetry using the NU method. In Section 3, we give our conclusions.

## 2. Bound State Solutions of the Combined Potential Model

The two radial coupled Dirac equations for the upper $F_{n\kappa}(r)$ and lower $G_{n\kappa}(r)$ spinor components can be expressed in the form [41]

$$\left(\frac{d}{dr} + \frac{\kappa}{r}\right) F_{n\kappa}(r) = \left(\frac{\tilde{\alpha}_\kappa}{\tilde{\gamma}_\kappa} - \Delta(r)\right) G_{n\kappa}(r), \quad (1a)$$

$$\left(\frac{d}{dr} - \frac{\kappa}{r}\right) G_{n\kappa}(r) = \left(\frac{\alpha_\kappa}{\gamma_\kappa} + \Sigma(r)\right) F_{n\kappa}(r), \quad (1b)$$

with

$$\tilde{\alpha}_\kappa = \tilde{\gamma}_\kappa (Mc^2 + E_{n\kappa}), \quad (2a)$$

$$\tilde{\gamma}_\kappa = \frac{1}{\hbar^2 c^2} (Mc^2 - E_{n\kappa} + C_{ps}), \quad (2b)$$

and

$$\alpha_\kappa = \gamma_\kappa (Mc^2 - E_{n\kappa}), \quad (3a)$$

$$\gamma_\kappa = \frac{1}{\hbar^2 c^2} (Mc^2 + E_{n\kappa} - C_s), \quad (3b)$$

where $\Delta(r) = V(r) - S(r)$ and the sum $\Sigma(r) = V(r) + S(r)$ are the difference and sum radial potentials, respectively. Also $c = 137.0359895$ is the velocity of light [35]. In the presence of the spin symmetry (i.e., $\Delta(r) = C_s =$ constant), one can eliminate $G_{n\kappa}(r)$ in (1a), with the aid of (1b), to obtain a second-order differential equation for the upper-spinor component as follows (for details see [41-45]):

$$\frac{d^2 F_{n\kappa}(r)}{dr^2} - \left(\frac{\kappa(\kappa+1)}{r^2} + \alpha_\kappa + \gamma_\kappa \Sigma(r)\right) F_{n\kappa}(r) = 0, \quad (4)$$

and the lower-spinor component can be obtained from (1a) as

$$G_{n\kappa}(r) = \frac{1}{\hbar^2 c^2 \gamma_\kappa} \left(\frac{d}{dr} + \frac{\kappa}{r}\right) F_{n\kappa}(r), \quad (5)$$

where $E_{n\kappa} \neq -Mc^2$ (only real positive energy states exist) when $C_s = 0$ (the exact spin symmetry case). On the other hand, under the pseudospin symmetry (i.e., $\Sigma(r) = C_{ps}$, where $C_{ps}$ a constant, one can eliminate $F_{n\kappa}(r)$ in (1b), with the aid of (1a), to obtain a second-order differential equation for the lower-spinor component as follows:

$$\frac{d^2 G_{n\kappa}(r)}{dr^2} - \left(\frac{\kappa(\kappa-1)}{r^2} + \tilde{\alpha}_\kappa - \tilde{\gamma}_\kappa \Delta(r)\right) G_{n\kappa}(r) = 0, \quad (6)$$

and the upper-spinor component $F_{n\kappa}(r)$ can be obtained from (1b) as

$$F_{n\kappa}(r) = \frac{1}{\hbar^2 c^2 \tilde{\gamma}_\kappa} \left(\frac{d}{dr} - \frac{\kappa}{r}\right) G_{n\kappa}(r), \quad (7)$$

where $E_{n\kappa} \neq Mc^2$ (only real negative energy states exist) when $C_{ps} = 0$ (for exact pseudospin symme-try). Thus, from the above equations, the energy eigenvalues depend on the quantum numbers $n$ and $\kappa$, and also the pseudo-orbital angular quantum number $\tilde{l}$ according to $\kappa(\kappa-1)=\tilde{l}(\tilde{l}+1)$, which implies that $j=\tilde{l}\pm 1/2$ are degenerate for $\tilde{l}\neq 0$. It is worthy to note that the reality and finiteness of our solutions demand that the upper and lower radial components are to satisfy the essential boundary conditions: $F_{n\kappa}(0)=G_{n\kappa}(0)=0$, and $F_{n\kappa}(\infty)=G_{n\kappa}(\infty)\to 0$.

We shall study the spin and pseudospin symmetric Dirac equation for the charged HO in a uniform electric field of strength $\varepsilon$ or the HO in an external dipole field. The corresponding scalar and vector components are taken to be the combined potential:

$$V(r)=U(r)+W(\varepsilon), \qquad (8)$$

where

$$U(r)=\frac{1}{2}M\omega_0^2 r^2, \qquad (9)$$

is the h.o. potential with the associated force given by

$$\vec{F}=-k_0\vec{r}, \qquad (10)$$

and

$$W(\varepsilon)=-q\varepsilon r, \qquad (11)$$

is the classical potential energy of a charged particle in a uniform external electric field of specific value $\varepsilon$ in the radial direction with $q$ being the charge of the particle. The combined potential (8) becomes $V(r')=M\omega_0^2 r'^2/2-q^2\varepsilon^2/2M\omega_0^2$ where $r'=r-q\varepsilon/M\omega_0^2$. In Fig. 1, we plot the ESP $V(r)$ versus distance $r$ with the choices of parameters (a) $M=1.0\,MeV$ and $\omega_0=1/2.4\,fm^{-1}$, (b) $M=1.5\,MeV$ and $\omega_0=1/2.4\,fm^{-1}$, (c) $M=1.0\,MeV$ and $\omega_0=1.0\,fm^{-1}$, and (d) $M=1.5\,MeV$ and $\omega_0=1.0\,fm^{-1}$ for several values of electric field strength: $\varepsilon=0, 0.5, 1.0$ and $2.0\,MeV\,fm^{-1}$.

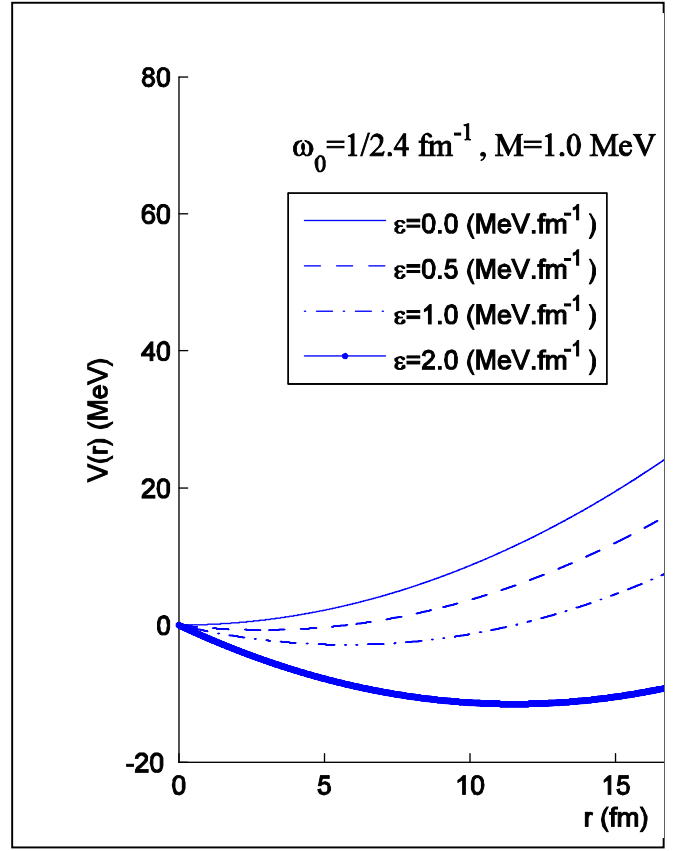

**Figure 1.** Variation of the $V(r)$ with distance $r$ in the presence and absence of electric field for the cases (a) $M=1.0\,MeV$ and $\omega_0=1/2.4\,fm^{-1}$,

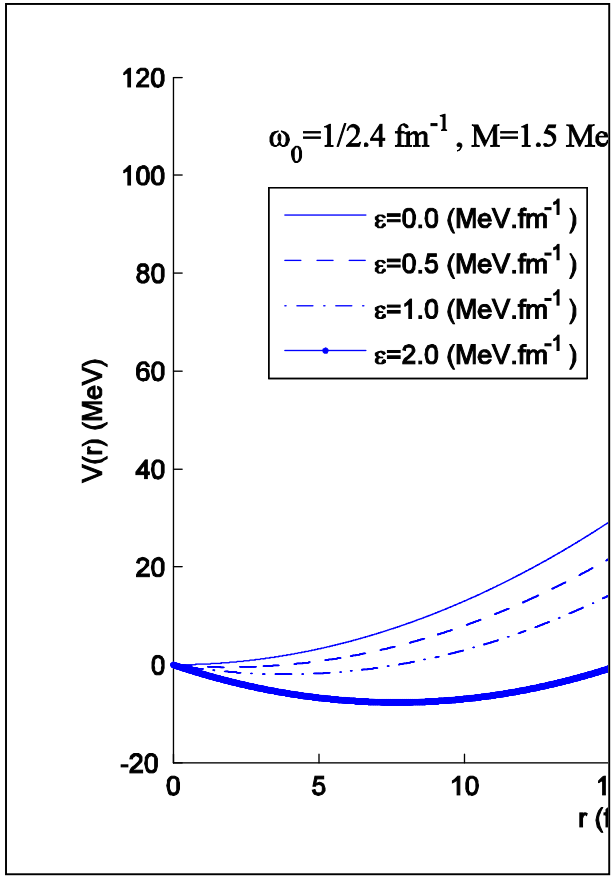

(b) $M = 1.5\, MeV$ and $\omega_0 = 1/2.4\, fm^{-1}$,

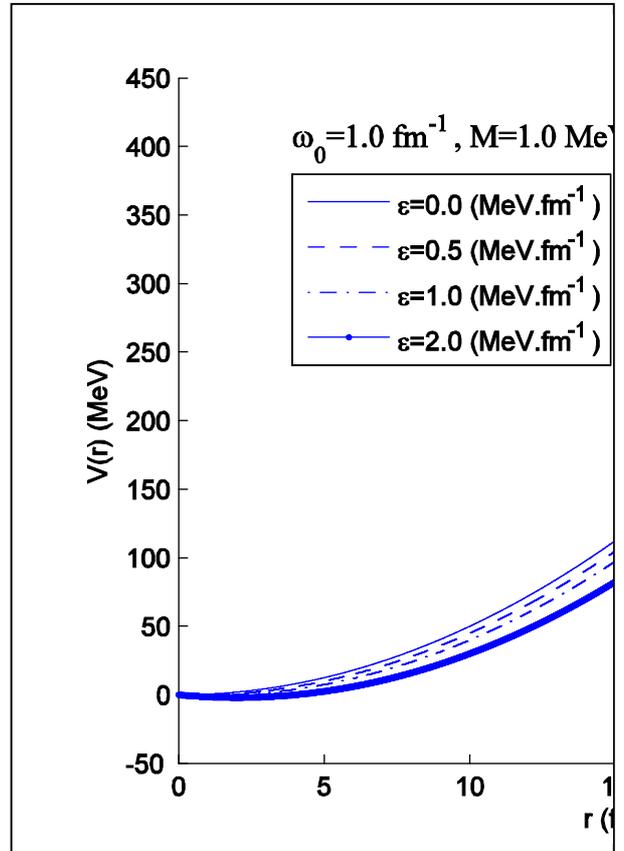

(c) $M = 1.0\, MeV$ and $\omega_0 = 1.0\, fm^{-1}$,

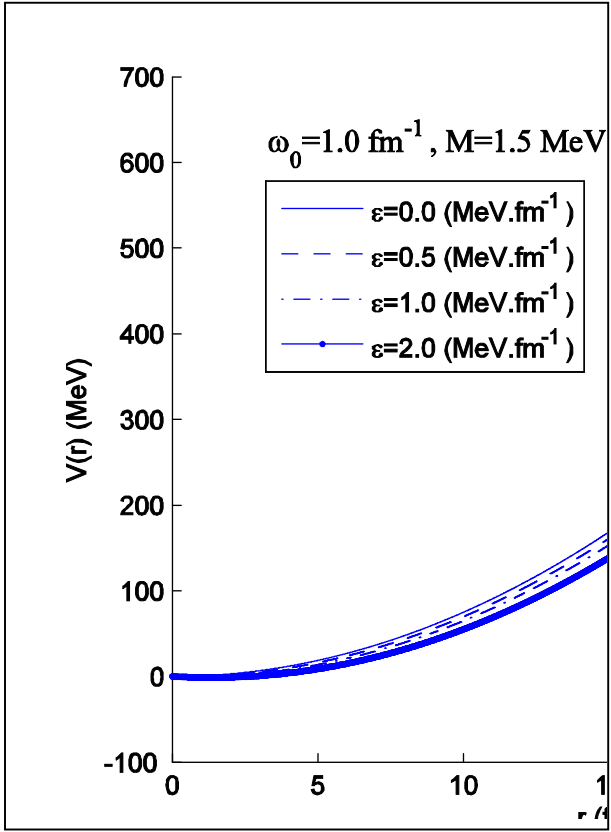

and (d) $M = 1.5\,MeV$ and $\omega_0 = 1.0\,fm^{-1}$.

It is clearly seen that the potential $V(r)$ has the harmonic-like behavior with frequency $\omega_0$, at least in the inner of nuclei, and at a radius $r_0 = 0$ describing the distance of well-bottom deviation from the center. Therefore, the study of spin and pseudospin partners splitting as a function of these parameters is meaningful and realistic enough to be applied to most nuclei at least qualitatively. It is known that, in certain isotope chains, as the mass number $A$ increases, the nuclear harmonic oscillator frequency $\omega_0$ decreases according to the $A^{1/3}$ law ( $\hbar\omega_0 = 41A^{-1/3}$ MeV ), which means that it is important to study the role of the parameter $\omega_0$ in spin and pseudospin symmetry [31]. In particular, the HO potential can provide fully bound states which are helpful to discuss the symmetry systematically.

### 2.1. The spin-symmetry solutions

Let us now study the potential model (8) in the context of spin symmetric Dirac equation (4). Therefore, it can be solved exactly for $\kappa = 0,-1$ because of the presence of spin-orbit centrifugal term. Under this symmetry, we take the sum potential in (4) as the combined potential, i.e.,

$$\Sigma(r) = \frac{1}{2}M\omega_0^2 r^2 - q\varepsilon r. \qquad (12)$$

We choose $\Sigma(r) = 2V(r) \to V(r)$ as stated in [41-45]*. The s-wave ( $\kappa = -1$, i.e., $l = 0$ ) case allows to rewrite (4) for the combined potential (12) as

$$\frac{d^2 F_{n\kappa}(r)}{dr^2} + \left(-\nu^2 r^2 + \beta r - \alpha_\kappa\right) F_{n\kappa}(r) = 0, \qquad (13)$$

where the constants are written as

$$\beta = q\varepsilon\gamma_\kappa \quad \text{and} \quad \nu = \sqrt{\frac{1}{2}M\omega_0^2 \gamma_\kappa}. \qquad (14)$$

To apply the NU method [5, 39], we need to compare (13) with (2) of [39] to obtain values for the parameters:

$$\tilde{\tau}(r) = 0, \sigma(r) = 0, \tilde{\sigma}(r) = -\nu^2 r^2 + \beta r - \alpha_\kappa. \qquad (15)$$

Inserting these values into (11) of [39], the selected forms of $\pi(r)$ and $k$ take the following particular values

$$\pi(r) = -\nu r + \frac{\beta}{2\nu}, \qquad (16)$$

and

$$k = \frac{\beta^2}{4\nu^2} - \alpha_\kappa, \qquad (17)$$

for discrete bound state solutions. According to the method, the following polynomial of degree one can be obtained from (6) of [39]:

$$\tau(r) = -2\nu r + \frac{\beta}{\nu} \quad \text{and} \quad \tau'(r) = -2\nu < 0, \qquad (18)$$

with prime denotes the derivative with respect to $r$. The parameters $\lambda$ and $\lambda_n$ in (7) and (10) of [39] take the simple forms as

$$\lambda = \frac{\beta^2}{4\nu^2} - \alpha_\kappa - \nu \quad \text{and} \quad \lambda_n = 2n\nu. \qquad (19)$$

Using the condition $\lambda = \lambda_n$, we obtain the transcendental energy equation for the charged particle confined by HO in an electric field of specific strength $\varepsilon$ as

$$\frac{\nu_{-1}}{\gamma_{-1}}(2n+1) = E_{n,-1} - Mc^2 + \frac{q^2 \varepsilon^2}{2M\omega_0^2}, n = 0,1,2,\cdots \qquad (20)$$

where at $\kappa = -1$, we have defined $\gamma_{-1} = \left(Mc^2 + E_{n,-1} - C_s\right)/\hbar^2 c^2 > 0$ and $\nu_{-1} = \sqrt{M\omega_0^2 \gamma_{-1}/2}$. We can compute the energy spectrum by choosing suitable parameters in the symmetric potential. Equation (20) shows the energy spectrum $E_{n,-1}$ dependence on $n$ and $C_s$ as well as on the parameters $\omega_0$ and $M$. If we choose $\varepsilon = 0$, (20) reduces to the one-dimensional energy spectrum of the relativistic HO:

$$\sqrt{\frac{E_{n,-1}+Mc^2}{2Mc^2}}\left(E_{n,-1}-Mc^2\right)=E_n^{\text{S-HO}}, \quad (21)$$

where $E_n^{\text{S-HO}} = (n+1/2)\hbar\omega$ is the well-known Schrödinger energy spectrum for the HO. The above result resembles the ones given in [22,33]. Hence, Dirac spectrum is composed of two sequences of discrete energy levels separated by the $Mc^2$ gap. This is Dirac oscillator [22,49-51] based on a construction of the Dirac equation which is ESP and in the NR limit gives the Schrödinger HO equation.

We use the parameter values of the HO potential $M=1.0\ GeV$ and $\omega_0=1.0\ GeV^2$ when $\varepsilon=0$ [52]. Hence, the numerical result for the s-wave energy spectrum of the bound state with relativistic corrections for the HO potential is: $E_0=1.4516059\ GeV$, $E_1=2.1880707\ GeV$, $E_0=2.8110575\ GeV$, and $E_3=3.3682575\ GeV$ for states $n=0,1,2$ and $3$, respectively.

We use (20) to compute some energy spectrum with relativistic corrections for several values of $n$. The computed exact spin symmetric ($C_s=0.0\ MeV$) energy spectrum is displayed in Table 1.

Also, for the combined potential we use the values $M=1.5\ MeV$ and $\omega_0=1/2.4\ fm^{-1}$. The strength of the electric field is set up at some arbitrary values of $\varepsilon=0,0.1,0.5,1.0$ and $2.0\ MeV.fm^{-1}$. Obviously, the energy levels are only positive under the spin symmetry limit when $\varepsilon=0-1.0\ MeV\ fm^{-1}$. However, as the field strength increases, i.e., $\varepsilon=2.0\ MeV\ fm^{-1}$, the results are noticed to become negative. The number of states in spectrum with negative values is finite (as $n$ increases, $E_{n,-1}\to 0$ and then energy eigenvalues flip their signs to positive values). In addition, the increase in the field strength, $\varepsilon > 2.0\ MeV\ fm^{-1}$, leads to no bound states. We conclude that the strength $\varepsilon$ has a maximum limit to provide real spectrum and hence must be adjusted carefully to produce real positive or negative values for the bound states.

The NR limit as a special case obtained when $C_s=0$ (exact spin symmetry) and employing appropriate parametric transformations: $\left(E_{n,-1}+Mc^2\right)/\hbar^2c^2\to 2M/\hbar^2$, and $E_{n,-1}-Mc^2\to E_n$ [28,42-45,53]. Therefore, the NR energy solution can be established for an electron confined in HO potential combined with an external electric field [see (8)] is

$$E_{n,l=0}^{\text{S-C}}=\hbar\omega_0\left(n+\frac{1}{2}\right)-\frac{q^2\varepsilon^2}{2M\omega_0^2},\ n=0,1,2,\cdots \quad (21)$$

leading to the ground-state energy spectrum formula:

$$E_{n=0}=\frac{\hbar^2}{2M}\left[\frac{M\omega_0}{\hbar}-\left(\frac{q\varepsilon}{\hbar\omega_0}\right)^2\right].$$

Obviously, we see from (21) that the entire spectrum of the harmonic oscillator (9) is shifted by the quantity $q^2\varepsilon^2/(2M\omega_0^2)$, this translation comes from the well-known fact that the electric field exerts a force on a charged particle. The modification is due only to the electric field. In the NR limit, the above result can be obtained directly from the well-known solution of the shifted HO in the absence of an electric field with change of variable $r'=r-q\varepsilon/M\omega_0^2$ and energy $E_n = E_n^{(\text{HO})}+c$, where the shifting energy $c=-q^2\varepsilon^2/2M\omega_0^2$. The above results are identical to the ones given in [11,18,19]. Taking $\varepsilon=0$, (21) is simply the NR HO solution (cf. e.g., [54]).

In Figure 2, (in units $\hbar=c=q=1$), we plot the energy spectrum in (21) versus the quantum number $n$ for specific values of $\varepsilon$ with the following choices: (a) $M=1.5\ MeV$, $\omega_0=1/2.4\ fm^{-1}$ and $q=1.0$ and (b) $M=1.5\ MeV$, $\omega_0=1.0\ fm^{-1}$ and $q=1.0$ for several values of electrical field strength, $\varepsilon=0,0.5,1.0$ and $2.0\ MeV.fm^{-1}$. As seen in Fig. 2a when the frequency is small and in the presence of stronger electrical field, energy states are shifted toward the negative energy part, i.e., for $\varepsilon=1.0\ MeV.fm^{-1}$, $n<6$ are strongly bound, however, when $\varepsilon=2.0\ MeV.fm^{-1}$, all states become strongly bound by the combined potential.

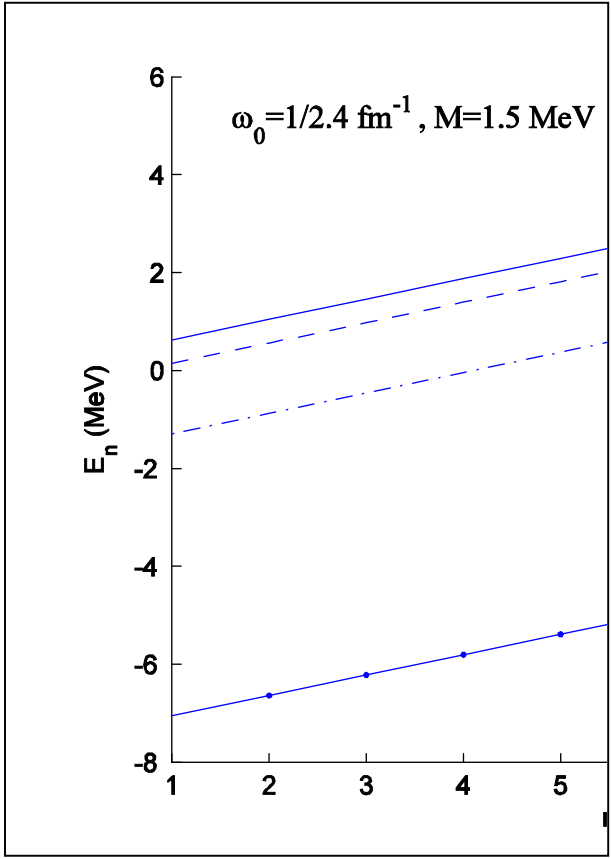
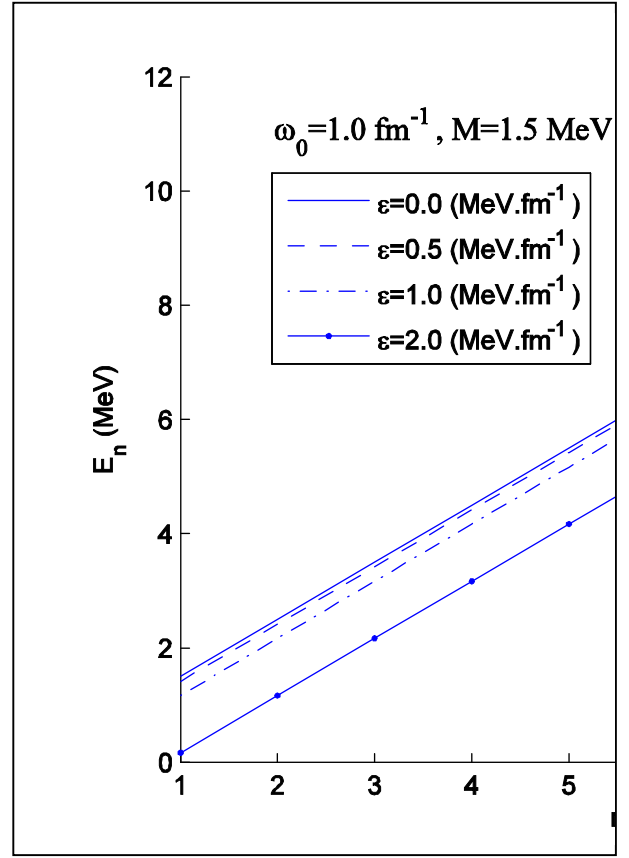

**Figure 2.** Energy spectrum $E_n$ versus $n$ in the presence and absence of uniform electric field for
(a) $M = 1.5\,MeV$ and $\omega_0 = 1/2.4\,fm^{-1}$, and (b) $M = 1.5\,MeV$ and $\omega_0 = 1/2.4\,fm^{-1}$.

On the other hand, in Fig. 2b, as frequency increases, the electrical field has no much effect in shifting energy levels; all states are in the positive energy part.

Next, we start the calculations of the corresponding wave functions. Consequently, both (4) and (9) of [39] give

$$\rho(r) = \exp\left[\varepsilon_2\left(\lambda^2 r - b\right)^2\right], \quad (22)$$

$$\phi(r) = \exp\left[\varepsilon_1\left(br - \frac{1}{2}\lambda^2 r^2\right)\right], \quad (23)$$

where

$$\lambda = \sqrt{\frac{M\omega_0}{\hbar}},\, b = \frac{q\varepsilon}{\hbar\omega_0},\, \varepsilon_1 = \sqrt{\frac{\hbar^2 \gamma_{-1}}{2M}},\, \varepsilon_2 = \frac{\hbar^2}{2M}\frac{\gamma_{-1}}{v}. \quad (24)$$

Hence, the first part of the wave function is

$$y_n(r) = H_n\left[\varepsilon_2\left(\lambda^2 r - b\right)^2\right] = L_n^{(0)}\left[\varepsilon_2\left(\lambda^2 r - b\right)^2\right], (25)$$

where $H_n(\beta z)$ is a nth degree Hermite polynomial, and $L_n^\alpha(r)$ is the associated Laguerre polynomial defined for the argument $r \in [0,\infty)$. The solution of (2) can be obtained by using (3) of [39] as

$$F_{n,-1} = N_n \exp\left[-\varepsilon_1\left(\frac{1}{2}\lambda^2 r^2 - br\right)\right] L_n\left[\varepsilon_2\left(\lambda^2 r - b\right)^2\right]. \quad (26)$$

In the NR limit, Eq. (26) becomes

$$R_n(r) = \left(\frac{\lambda^2}{\pi}\right)^{1/4} \frac{1}{\sqrt{2^n n!}} \exp\left[-\frac{\lambda^2}{2}\left(r - \frac{q\varepsilon}{M\omega_0^2}\right)^2\right]$$

$$\times H_n\left(\lambda\left(r-\frac{q\varepsilon}{M\omega_0^2}\right)\right), \quad (27)$$

where $\langle r \rangle = -q\varepsilon/M\omega_0^2$, for all eigenstates $r$ and for the ground state it is given by $R_0(r) = (\lambda^2/\pi)^{1/4} \exp\left[-\lambda^2(r-q\varepsilon/M\omega_0^2)^2/2\right]$, where $H_0(\lambda(r-q\varepsilon/M\omega_0^2)) = 1$.

As seen, the modification in (26) is essentially produced from the external electric field. Furthermore, the associated lower-spinor component $G_{n,-1}(r)$ satisfying Eq. (5) is taking the form

$$G_{n,-1} = N_n' \exp\left[-\varepsilon_1(\lambda^2 r^2/2 - br)\right]$$

$$\times\left\{\left[\varepsilon_1(b-\lambda^2 r) + \frac{\kappa}{r}\right]L_n^{(0)}\left[\varepsilon_2(\lambda^2 r - b)^2\right]\right.$$

$$\left. + 2\lambda^2 \varepsilon_2(\lambda^2 r - b)L_n^{(1)}\left[\varepsilon_2(\lambda^2 r - b)^2\right]\right\}, \quad (28)$$

where $N_n' = d_0 N_n$ with $d_0 = 1/\hbar^2 c^2 \gamma_{-1}$ is counted as a new normalization constant. It is worth to mention that $E_{n,-1} \neq -Mc^2$ under the exact spin symmetry $C_s = 0$.

## 2.2. The Pseudospin Symmetric Solutions

The exact pseudospin symmetry occurs when $S(r) \approx -V(r)$ [16,23-25,40] and the quality of the pseudospin approximation in real nuclei is connected with the competition between the pseudo-centrifugal barrier and the pseudospin-orbital potential [31]. Here $\Delta(r)$ in Eq. (6) is taken to be the same as the combined potential (8), that is,

$$\Delta(r) = \frac{1}{2}M\omega_0^2 r^2 - q\varepsilon r. \quad (29)$$

Thus, for the s-wave $(\kappa = 1)$, Eq. (6) can be rewritten as

$$\frac{d^2 G_{n\kappa}(r)}{dr^2} + \left(-\tilde{v}^2 r^2 + \tilde{\beta} r - \tilde{\alpha}_\kappa\right)G_{n\kappa}(r) = 0, \kappa = 1 \quad (30)$$

where

$$\tilde{\beta} = q\varepsilon\tilde{\gamma}_\kappa \quad \text{and} \quad \tilde{v} = \sqrt{\frac{1}{2}M\omega_0^2 \tilde{\gamma}_\kappa}. \quad (31)$$

To avoid repetition in the solution of Eq. (30), a first inspection for the relationship between the present set of parameters $(\tilde{\alpha}_\kappa, \tilde{\beta}, \tilde{v})$ and the previous set $(\alpha_\kappa, \beta, v)$ tells us that the energy solution for pseudospin symmetry can be easily obtained directly from the spin symmetry energy solutions by performing the transformation changes [42-45,53,55]:

$$F_{n,-1}(r) \leftrightarrow G_{n,1}(r), E_{n,-1} \leftrightarrow -E_{n,1}, C_s \leftrightarrow -C_{ps},$$

$$V(r) \leftrightarrow -V(r) \text{ (or } \alpha_\kappa \to \tilde{\alpha}_\kappa, v^2 \to -\tilde{v}^2, \beta \to -\tilde{\beta}), \quad (32)$$

or alternatively, the essential parameters given by
$\tilde{\tau} = 0, \sigma = 1, \tilde{\sigma} = \tilde{v}^2 r^2 - \tilde{\beta} r - \tilde{\alpha}$,

$$\pi = -i(\tilde{v}r - \tilde{\beta}/2\tilde{v}), k = -(\tilde{\beta}^2/4\tilde{v}^2 + \tilde{\alpha}),$$

$$\tau = -i(2\tilde{v}r - \tilde{\beta}/\tilde{v}), \tau' = -2i\tilde{v},$$

$$\lambda = -(\tilde{\beta}^2/4\tilde{v}^2 + \tilde{\alpha} + i\tilde{v}) \text{ and } \lambda_n = -2i\tilde{v}n.$$

are used to obtain the following transcendental energy equation:

$$2n+1 = -i\left(\frac{q^2\varepsilon^2}{2M\omega_0^2} + E_{n,1} + Mc^2\right)\frac{\tilde{\gamma}_1}{\tilde{v}_1}, n=0,1,2,\cdots, (33)$$

with

$$\tilde{v}_1 = \sqrt{\frac{1}{2}M\omega_0^2 \tilde{\gamma}_1}, \tilde{\gamma}_1 = \frac{1}{\hbar^2 c^2}(Mc^2 - E_{n,1} + C_{ps}) > 0$$

where $E_{n,1} > Mc^2 + C_{ps}$ is the restriction condition for the discrete bound states. Hence, Eq. (33) is identical to Eq. (62) of [23] obtained for the HO potential if the electric field strength $\varepsilon$ is set to zero.

The NR limit of Eq. (33), when $C_{ps} = 0$, becomes

$$E_{n,1} = \frac{\hbar^2 \omega_0^2}{2M}\left(n+\frac{1}{2}\right)^2 \left[1+\left(\frac{q\varepsilon}{2M\omega_0}\right)^2\right]^{-2}, (34)$$

where $\omega_0$ is the classical frequency for small harmonic vibrations [56]. The right hand side of (34) is always positive. Therefore, there are only positive energies in the NR limit for the HO potential [28,35]. Hence, using (33), we can compute the energies by choosing suitable parameters in the pseudo symmetric limit. The energies $E_{n,1}$ are dependent on $n$ and $C_{ps}$ as well as on the parameters $\omega_0$ and $M$. We also compute the energy spectrum of the bound state system with relativistic corrections for several values of $n$ with parameter values of the potential $M = 1.5\, MeV$ and $\omega_0 = 1/2.4\, fm^{-1}$ (for two constants $C_{ps} = -10.3\, MeV$ and $-11.5\, MeV$ as represented in Table 2. The external electrical field is set up at some values $\varepsilon = 0, 0.1, 0.5, 1.0$ and $1.5\, MeV\, fm^{-1}$.

We see that there are only negative energy bound state solutions in the pseudospin symmetry limit when the strength of external electric fields are $\varepsilon = 0 - 1.81\, MeV\, fm^{-1}$. Nevertheless, as the field strength increases, i.e., $\varepsilon > 1.90\, MeV\, fm^{-1}$, the energies are noticed to become complex for all states. We conclude that when the external electric field $\varepsilon$ strength exceeds the ionization limit then it provides imaginary spectrum (no bound states). Hence the strength of the electric field must be adjusted carefully to generate real positive/negative bound states. Also, the parameters $M$ and $C_{ps}$ must be adjusted properly for real solutions of the transcendental energy equation (33).

Finally, we calculate the lower-spinor wave function which is the solution of (6) as

$$G_{n,1}(r) = \widetilde{N}_n \exp\left[i\hbar c \varepsilon_1'\left(-br + \frac{\lambda^2}{2}r^2\right)\right] \times H_n\left[-i\hbar c \varepsilon_2'(\lambda^2 r - b)^2\right], \quad (35)$$

where $\varepsilon_1' = \sqrt{\widetilde{\gamma}_1/2M}\hbar$ and $\varepsilon_2' = \hbar^2 \widetilde{\gamma}_1/2M\widetilde{v}_1$.

## 3. Conclusions

The exact s-wave Dirac bound states (energy spectra and wave functions) of the potential (8) in the presence of the spin symmetry and pseudospin symmetry are obtained in closed form using the NU method. The wave functions are expressed in terms of the orthogonal Laguerre polynomials. For the exact spin symmetry (i.e., $C_s = 0$), the relativistic solution can be readily reduced to the NR one by an appropriate mapping transformations. The presence of an external uniform electric field creates a shift at the energy spectrum and a translation on the wave functions for the HO. In case if the description of diatomic vibration motion is NR, the relativistic model used seems quite justified since it can be easily reduced to the NR limits [18,19]. As numerical example, we take a set of physical parameter values to determine the bound state energy eigenvalues as shown in Tables 1 and 2 for the spin and pseudospin symmetry cases, respectively. It is worth to mention that the strength of the applied electric field $\varepsilon$ needs to be adjusted properly to provide us bound state energy spectrum for certain values of parameters $\omega_0$, $M$, $C_s$ and $C_{ps}$.

## 4. Acknowledgements

The author thanks the two kind referees for their enlightening suggestions.

## 5. Appendix A: The Solution of Cubic Energy Equation

Let us solve the general cubic energy equation:

$$AE_{n,-1}^3 + BE_{n,-1}^2 + CE_{n,-1} + D = 0; \quad A \neq 0, \quad (A1)$$

where

$$A = 1, B = M_s + 2g(\varepsilon), C = g(\varepsilon)[g(\varepsilon) + 2M_s],$$
$$D = g^2(\varepsilon)M_s - 2M\omega_0^2(n+1/2)^2,$$
$$g(\varepsilon) = q^2\varepsilon^2/2M\omega_0^2 - M, \quad M_s = M - C_s, \quad (A2)$$

are real or complex numbers. We want to reduce the cubic equation (A1) to a 'depressed' cubic (i.e., the quadratic term disappears) via $E_{n,-1} = y - B/3A$, to obtain

$$y^3 + dy = -e, \quad d = C - B^2/(3A^2),$$
$$e = D + B(2B^2 - 9AC)/(27A^2), \quad (A3)$$

where

$$d = -[M_s - g(\varepsilon)]^2/3, \quad \text{and}$$
$$e = 2[M_s - g(\varepsilon)]^3/27 - 2M\omega_0^2(n+1/2)^2. \quad (A4)$$

Further, the substitution $y = z - d/3z$ in (A3), we get

$$z^3 - (d/3z)^3 + e = 0, \quad (A5)$$

and multiplying throughout the above equation by $z^3$, we obtain the quadratic form

$$(z^3)^2 + e(z^3) + p = 0, \quad p = -(d/3)^3. \quad (A6)$$

This equation can be easily solved for real bound states:

$$z = \left(-\frac{e}{2} \pm \frac{1}{2}\sqrt{e^2 - 4p}\right)^{1/3}, \quad e^2 \geq 4p. \quad (A7)$$

Thus, the energy in the presence of electric field reads

$$E_{n,-1}(\varepsilon) = \left(-\frac{e}{2} \pm \frac{1}{2}\sqrt{e^2 - 4p}\right)^{1/3}$$
$$-\frac{d}{3}\left(-\frac{e}{2} \pm \frac{1}{2}\sqrt{e^2 - 4p}\right)^{-1/3} - \frac{B}{3}, \quad (A8)$$

and also in the absence of electric field,

$$E_{n,-1}(\varepsilon = 0) = \left(-\frac{u}{2} \pm \frac{1}{2}\sqrt{u^2 - 4\left(\frac{M_s}{3}\right)^6}\right)^{1/3}$$
$$+ \frac{M_s^2}{9}\left(-\frac{u}{2} \pm \frac{1}{2}\sqrt{u^2 - 4\left(\frac{M_s}{3}\right)^6}\right)^{-1/3} - \frac{M_s}{3}, \quad (A9)$$

where $u = 3M_s^3/27 - 2M\omega_0^2(n+1/2)^2$.

[1] O.L. de Lange and R.E. Raab, "Operator Methods in Quantum Mechanics," Clarendon: Oxford, 1991.

[2] G. Arfken, "Mathematical Methods for Physicists," Academic Press: San Diego, CA, 1995.

[3] C.V. Sukumar, "Supersymmetric Quantum Mechanics of One-Dimensional Systems," J. Phys. A: Math. Gen., Vol. 18, 1985, pp. 2917-2936.

[4] F. Cooper, A. Khare and U. Sukhatme, "Supersymmetry in Quantum Mechanics," World Scientific, Singapore, 2001.

[5] A.F. Nikiforov and V.B. Uvarov, "Special Functions of Mathematical Physics," Birkhäuser: Basel, 1988.

[6] O. Bayrak, I. Boztosun and H. Ciftci, "Exact Analytical

Solutions to the Kratzer Potential by the Asymptotic Iteration Method," Int. J. Quantum Chem., Vol. 107, 2007, pp. 540-544.

[7] W.C. Qiang and S.H. Dong, "Arbitrary $l$-State Solutions of the Rotating Morse Potential Through the Exact Quantization Rule Method," Phys. Lett. A, Vol. 363, 2007, pp. 169-176.

[8] I. Nasser, M.S. Abdelmonem, H. Bahlouli and A.D. Alhaidari, "The Rotating Morse Potential Model for diatomic Molecules in the Tridiagonal J-Matrix Representation: 1. Bound States," J. Phys. B: At. Mol. Opt. Phys., Vol. 40, 2007, pp. 4245-4258.

[9] R.R. Freeman, N.P. Economou, G.C. Bjorklund and K.T. Lu, "Observation of Electric Field Induced Resonances Above the Ionization Limit in a One Electron Atom," Phys. Rev. Lett., Vol. 41, No. 21, 1978, pp. 1463-1467.

[10] J.M. Jauch, "The Hyperfine Structure and the Stark Effectof the Ammonia Inversion Spectrum," Phys. Rev., Vol. 72, 1947, pp. 715-723.

[11] J.J. Peña, M.A. Romero-Romo, J. Morales and J.L. López-Bonilla, "Exactly Solvable Combined Potentials and their Isospectral Partners From SUSY," Int. J. Quantum Chem., Vol. 105, 2005, pp. 731-739.

[12] E. Witten, "Super-Symmetry and Other Scenarios," Int. J. Mod. Phys. A, Vol. 19, No. 8, 2004, pp. 1259-1264.

[13] H.T. Davis, "Introduction to Nonlinear Differential and Integral Equations," Dover: New York, 1962.

[14] S.M. Ikhdair, "Rotation and Vibration of Diatomic Molecule in the Spatially Dependent Mass Schrödinger Equation With Generalized $q$-Deformed Morse Potential," Chem. Phys., Vol. 361, No. 1-2, 2009, pp. 9-17.

[15] S.M. Ikhdair, "An Improved Approximation Scheme for the Centrifugal Term and the Hulthen Potential," Eur. Phys. J. A, Vol. 39, 2009, pp. 307-314.

[16] S.M. Ikhdair and R. Sever, "Any $l$-State Improved Quasi-Exact Analytical Solutions of the Spatially Dependent Mass Klein—Gordon Equation for the scalar and Vector Hulthen Potential," Phys. Scr., Vol. 79, 2009, pp. 035002 (12 pages).

[17] S.M. Ikhdair and R. Sever, "Exact Quantization Rule to the Kratzer-Type Potentials: An Application to the Diatomic Molecules," J. Math. Chem., Vol. 45, No. 4, 2009, pp. 1137-1152.

[18] Sérgio Azevedo, "Harmonic Oscillator in a Space With a Linear Topological Defect," Phys. Lett. A, Vol. 288, 2001, pp. 33-36.

[19] Sérgio Azevedo, "Influence of the Electric Field on a Particle in a Space with a Disclination," Int. J. Quantum Chem., Vol. 101, 2005, pp. 127-130.

[20] Q.W. Chao, "Bound States of the Klein-Gordon and Dirac Equations for Scalar and Vector Harmonic Oscillator Potentials," Chin. Phys., Vol. 11, 2002, pp. 757-759.

[21] T.-S. Chen, H.-F. Lü, J. Meng, S.-Q. Zhang and S.-G. Zhou, "Pseudospin Symmetry in Relativistic Framework With Harmonic Oscillator Potential and Woods-Saxon Potential," Chin. Phys. Lett., Vol. 20, 2003, pp. 358-361.

[22] V.I. Kukulin, G. Loyola and M. Moshinsky, "A Dirac Equation With an Oscillator Potential and Spin-Orbit Coupling," Phys. Lett. A, Vol. 158, 1991, pp. 19-22.

[23] J.N. Ginocchio, "Relativistic Harmonic Oscillator With Spin Symmetry," Phys. Rev. C, Vol. 69, 2004, pp. 034318 (8 pages).

[24] J.N. Ginocchio, "Pseudospin As a Relativistic Symmetry," Phys. Rev. Lett., Vol. 78, 1997, pp. 436-459.

[25] J.N. Ginocchio, "Relativistic Symmetries in Nuclei and Hadrons," Phys. Rep., Vol. 414, 2005, pp. 165-261.

[26] P. Alberto, M. Fiolhais, M. Malheiro, A. Delfino and M. Chiapparini, "Isospin Asymmetry in the PseudospinDynamical Symmetry," Phys. Rev. Lett., Vol. 86, 2001, pp. 5015-5018.

[27] P. Alberto, M. Fiolhais, M. Malheiro, A. Delfino and M. Chiapparini, "Pseudospin Symmetry As a Relativistic Dynamical Symmetry in the Nucleus," Phys. Rev. C, Vol. 65, 2002, pp. 034307 (9 pages).

[28] R. Lisboa, M. Malheiro, A.S. de Castro, P. Alberto and M. Fiolhais, "Pseudospin Symmetry and the Relativistic Harmonic Oscillator," Phys. Rev. C, Vol. 69, 2004, 024319 (15 pages).

[29] R. Lisboa, M. Malheiro, A.S. de Castro, P. Alberto and M. Fiolhais, "Perturbative Breaking of the Pseudospin Symmetry in the Relativistic Harmonic Oscillator," Int. J. Mod. Phys. D, Vol. 13, 2004, pp. 1447-1451.

[30] J.N. Ginocchio, "A Relativistic Symmetry in Nuclei," Phys. Rep., Vol. 315, 1999, pp. 231-240.

[31] J.-Y. Guo, X.-Z. Fang and F.-X. Xu, "Pseudospin Symmetry in the relativistic Harmonic Oscillator," Nuclear Phys. A, Vol. 757, 2005, pp. 411-421.

[32] A.S. de Castro, P. Alberto, R. Lisboa and M. Malheiro, "Rotating Pseudospin and Spin Symmetries Through Charge-Conjugation and Chiral Transformations: the case of the Relativistic Harmonic Oscillator," Phys. Rev. C, Vol. 73, 2006, pp. 054309 (13 pages).

[33] H. Akçay and C. Tezcan, "Exact Solutions of the Dirac EquationWith Harmonic Oscillator Potential Including a Coulomb-Like Tensor Potential," Int. J. Mod. Phys. C, Vol. 20, No. 6, 2009, pp. 931-940.

[34] Y. Zhou and J.-Y. Guo, "The Relativistic Bound States for a New Ring-Shaped Harmonic Oscillator," Chin. Phys. B, Vol. 17, No. 2, 2008, pp. 380-384.

[35] O. Aydoğdu and R. Sever, "Solution of the Dirac Equation for Pseudoharmonic Potential by Using the Nikiforov-Uvarov Method," Phys. Scr., Vol. 80, 2009, pp. 015001 (6 pages).

[36] S.M. Ikhdair and R. Sever, "Improved Analytical Approximation to Arbitrary $l$-State Solutions of the Schrödinger equation for the Hyperbolical Potentials," Ann. Phys. (Berlin), Vol. 18, 2009, pp. 747-758.

[37] S.M. Ikhdair, "On the Bound-State Solutions of the Manning-Rosen Potential Including an Improved Approximation to the Orbital Centrifugal Term," Phys. Scr., Vol. 83, 2011, pp. 015010 (7 pages).

[38] S.M. Ikhdair, "Exactly Solvable Effective Mass $D$ Dimensional Schrödinger equation for Pseudoharmonic and

Modified Kratzer potentials," Int. J. Mod. Phys. C, No.3, Vol. 20, 2009, pp. 361-372.

[39] S.M. Ikhdair, "Exact Klein-Gordon Equation with Spatially Dependent Masses for Unequal Scalar-Vector Coulomb-Like Potentials," Eur. Phys. J. A, Vol. 40, No. 2, 2009, pp. 143-149.

[40] J. Meng, K. Sugawara-Tanabe, S. Yamaji and A. Arima, "Pseudospin Symmetry in Zr and Sn Isotopes from the Proton Drip Line to the Neutron Drip Line," Phys. Rev. C, Vol. 59, 1999, pp. 154.

[41] W. Greiner, "Relativistic Quantum Mechanics," Springer, Verlag, 1981.

[42] S.M. Ikhdair, "Approximate Solutions of the Dirac Equation for the Rosen-Morse Potential Including Spin-Orbit Centrifugal Term," J. Math. Phys., Vol. 51, 2010, pp. 023525 (16 pages); S.M. Ikhdair, "An Approximate $\kappa$ State Solutions of the Dirac Equation for the Generalized Morse potential Under Spin and Pseudospin Symmetry," J. Math. Phys., Vol. 52, 2011, pp. 052303 (22 pages).

[43] S.M. Ikhdair and R. Sever, "Approximate Analytical Solutions of the Generalized Woods-Saxon Potential ıncluding Spin-Orbit Coupling term and Spin symmetry," Cent. Eur. J. Phys., Vol. 8, No. 4, 2010, pp. 652-666.

[44] S.M. Ikhdair and R. Sever, "Approximate Bound state Solutions of Dirac Equation With Hulthen Potential ıncluding Coulomb-Like tensor potentials," Appl. Math. Comput., Vol. 216, 2010, pp. 911-923; S.M. Ikhdair and R. Sever, "Solutions of the Spatially-Dependent Mass dirac equation With the Spin and Pseudospin Symmetry for the Coulomb-Like Potential," Appl. Math. Comput., Vol. 216, 2010, pp. 545-555.

[45] S.M. Ikhdair and R. Sever, "Two Approximation Schemes to the Bound-States of the Dirac-Hulthen Problem," J. Phys. A: Math. and Theor., Vol. 44, 2011, pp. 355301 (29 pages).

[46] J. Karwowski, G. Pestka, M. Stanke and F.E. Harris, "Representation of the Dirac Equation and the VariationalPrinciple," Int. J. Quantum Chem., Vol. 106, 2006, pp. 3129-3139.

[47] M. Stanke, J. Karwowski and H. Tatewaki, "Kinetically Balanced Dirac Equation:Properties and Application," Mol. Phys., Vol. 104, 2004, pp. 2085-2092.

[48] S.K. Bose, A. Schulze-Halberg and M. Singh, "New Exact Solutions of the Dirac Equation," Phys. Lett. A, Vol. 287, 2001, pp. 321-324.

[49] M. Moshinsky, "The Harmonic Oscillator in Modern Physics: From Atoms to Quarks," Gordon and Breach, New York, pp. 29, 1969.

[50] M. Moshinsky and A. delSol Mesa, "The Dirac Oscillator of Arbitrary Spin," J. Phys. A: Math. Gen., Vol. 29, 1996, pp. 4217-4236.

[51] R.M. Mir-Kasimov, "$SU_q(1,1)$ and the Relativistic Oscillator," J. Phys. A: Math. Gen., Vol. 24, 1991, pp. 4283-4302.

[52] W. Lucha, H. Rupprecht and F.F. Schöberl, "Significance of Relativistic Wave Equations for Bound States," Phys. Rev. D, Vol. 46, 1992, pp. 1088-1095.

[53] A.D. Alhaidari, H. Bahlouli and A. Al-Hasan, "Dirac and Klein-Gordon Equations With Equal Scalar and Vector Potentials," Phys. Lett. A, Vol. 349, 2006, pp. 87-97.

[54] L.I. Schiff, "Quantum Mechanics," 3rd Edition, McGraw-Hill, New York, 1968.

[55] C. Berkdemir and Y.-F. Cheng, "On the Exact Solutions of the Dirac Equation With a Novel Angle-Dependent Potential," Phys. Scr., Vol. 79, 2009, 035003.

[56] S. Flügge, "Practical Quantum Mechanics I," Berlin: Springer, 1971.

**Table 1. Energy levels (in relativistic units $\hbar = c = q = 1$) for different quantum numbers $n$.**

| $n$ | $\varepsilon = 0.0$ | $\varepsilon = 0.1$ | $\varepsilon = 0.5$ | $\varepsilon = 1.0$ | $\varepsilon = 1.5$ | $\varepsilon = 2.0$ |
|---|---|---|---|---|---|---|
| $C_s = 0.0\ MeV^a$ | | | | | | |
| 0 | 0.271140 | 0.253314 | -0.167413 | -1.22805 | -1.48381 | -1.49659 |
| 1 | 0.725628 | 0.709133 | 0.321988 | -0.705506 | -1.36573 | -1.46962 |
| 2 | 1.11559 | 1.09978 | 0.728580 | -0.283911 | -1.17160 | -1.41701 |
| 3 | 1.46654 | 1.45115 | 1.08963 | 0.085816 | -0.941148 | -1.34121 |
| 4 | 1.79036 | 1.77528 | 1.42039 | 0.422329 | -0.696652 | -1.24528 |
| 5 | 2.09380 | 2.07894 | 1.72893 | 0.735034 | -0.448725 | -1.13249 |
| 6 | 2.38113 | 2.36645 | 2.02022 | 1.02949 | -0.202223 | -1.00597 |
| 7 | 2.65528 | 2.64074 | 2.29754 | 1.30932 | 0.040681 | -0.868546 |
| 8 | 2.91835 | 2.90393 | 2.56322 | 1.57705 | 0.279078 | -0.722603 |
| 9 | 3.17194 | 3.15761 | 2.81899 | 1.83453 | 0.512668 | -0.570129 |
| 10 | 3.41725 | 3.40301 | 3.06619 | 2.08318 | 0.741446 | -0.412735 |

| $C_s = -5.0\, MeV^a$ | | | | | | |
|---|---|---|---|---|---|---|
| 0 | 0.140034 | 0.121035 | -0.334675 | -1.75436 | -4.08767 | -6.41822 |
| 1 | 0.411762 | 0.393118 | -0.053633 | -1.43881 | -3.67584 | -6.05578 |
| 2 | 0.673628 | 0.65529 | 0.216190 | -1.14065 | -3.30985 | -5.68341 |
| 3 | 0.926861 | 0.908791 | 0.476322 | -0.856711 | -2.97470 | -5.33739 |
| 4 | 1.17245 | 1.15462 | 0.727960 | -0.584715 | -2.66224 | -5.01500 |
| 5 | 1.41121 | 1.39358 | 0.972082 | -0.322940 | -2.36745 | -4.71176 |
| 6 | 1.64380 | 1.62636 | 1.20947 | -0.070054 | -2.08697 | -4.42415 |
| 7 | 1.87080 | 1.85353 | 1.44079 | 0.175004 | -1.81842 | -4.14953 |
| 8 | 2.09269 | 2.07557 | 1.66658 | 0.413093 | -1.56002 | -3.88591 |
| 9 | 2.30987 | 2.29290 | 1.88734 | 0.644922 | -1.31041 | -3.63177 |
| 10 | 2.52272 | 2.50588 | 2.10346 | 0.871085 | -1.06853 | -3.38590 |
| $C_s = 5.0\, MeV^a$ | | | | | | |
| 0 | 3.51057 | 3.51045 | 3.50819 | 3.50443 | 3.50213 | 3.50104 |
| 1 | 3.59088 | 3.58996 | 3.57140 | 3.53932 | 3.51907 | 3.50936 |
| 2 | 3.73353 | 3.73141 | 3.68743 | 3.60658 | 3.55252 | 3.52592 |
| 3 | 3.91604 | 3.91270 | 3.84162 | 3.70187 | 3.60167 | 3.55059 |
| 4 | 4.12103 | 4.11660 | 4.02068 | 3.82010 | 3.66540 | 3.58314 |
| 5 | 4.33744 | 4.33211 | 4.21481 | 3.95627 | 3.74238 | 3.62331 |
| 6 | 4.55882 | 4.55275 | 4.41745 | 4.10599 | 3.83119 | 3.67079 |
| 7 | 4.78145 | 4.77477 | 4.62442 | 4.26568 | 3.9304 | 3.72522 |
| 8 | 5.00325 | 4.99606 | 4.83305 | 4.43250 | 4.0386 | 3.78622 |
| 9 | 5.22305 | 5.21543 | 5.04170 | 4.60428 | 4.15451 | 3.85337 |
| 10 | 5.44020 | 5.43222 | 5.24932 | 4.77940 | 4.27694 | 3.92628 |

a. For spin symmetry case, we used the procedures explained in Appendix A.

**Table 2.** Energy levels (in relativistic units $\hbar = c = q = 1$) for different quantum numbers $n$.

| $C_{ps} = -10.3\, MeV^b$ | | | | | | $C_{ps} = -11.5\, MeV^b$ | | | | |
|---|---|---|---|---|---|---|---|---|---|---|
| $n$ | $\varepsilon = 0.0$ | $\varepsilon = 0.1$ | $\varepsilon = 0.5$ | $\varepsilon = 1.0$ | $\varepsilon = 1.5$ | $\varepsilon = 0.0$ | $\varepsilon = 0.1$ | $\varepsilon = 0.5$ | $\varepsilon = 1.0$ | $\varepsilon = 1.5$ |
| 0 | -1.635 | -1.654 | -2.120 | -3.578 | -6.037 | -1.625 | -1.644 | -2.108 | -3.562 | -6.000 |
| 1 | -1.912 | -1.932 | -2.408 | -3.910 | -6.540 | -1.880 | -1.900 | -2.372 | -3.857 | -6.390 |
| 2 | -2.202 | -2.223 | -2.711 | -4.267 | -7.287 | -2.144 | -2.164 | -2.645 | -1.494 | -6.834 |
| 3 | -2.507 | -2.528 | -3.032 | -4.662 | | -2.417 | -2.438 | -2.930 | -4.497 | -7.381 |
| 4 | -2.829 | -2.851 | -3.374 | -5.111 | | -2.702 | -2.723 | -3.228 | -4.851 | -8.343 |
| 5 | -3.173 | -3.196 | -3.746 | -5.660 | | -3.000 | -3.022 | -3.542 | -5.239 | |
| 6 | -3.547 | -3.571 | -4.157 | -6.541 | | -3.314 | -3.336 | -3.876 | -5.676 | |
| 7 | -3.960 | -3.986 | -4.631 | | | -3.648 | -3.671 | -4.234 | -6.195 | |
| 8 | -4.437 | -4.466 | -5.224 | | | -4.005 | -4.030 | -4.626 | -6.909 | |
| 9 | -5.032 | -5.068 | | | | -4.396 | -4.422 | -5.067 | | |
| 10 | -6.195 | | | | | -4.834 | -4.862 | -5.587 | | |

b. Pseudospin symmetric case.